\documentclass[a4paper,11pt]{article}
\pdfoutput=0 

\usepackage{jheppub} 

\usepackage[T1]{fontenc} 

\usepackage{graphicx}

\newcommand*{\be}
{\begin{equation}}
\newcommand*{\ee}
{\end{equation}}
\newcommand*{\beq}
{\begin{eqnarray}}
\newcommand*{\eeq}
{\end{eqnarray}}
\newcommand*{\nn}
{\nonumber}
\newcommand*{\LL}
{{\cal L}}
\newcommand*{\hPhi}
{{\hat \Phi}}

\newcommand*{\hphi}
{{\hat \phi}}
\newcommand*{\dd}
{{\displaystyle}}

\title{\boldmath Variational approach to thermal masses in compactified models }

\author[a]{Daniele Dominici}
\author[b]{Itzhak Roditi}

\affiliation[a]{Dipartimento di Fisica e Astronomia Universit\`a di Firenze and INFN, Sezione di Firenze, 50019 Sesto F., Italy}
\affiliation[b]{Centro Brasileiro de Pesquisas F\'{\i}sicas - CBPF/MCT, 22290-180, Rio de Janeiro, RJ, Brazil}

\emailAdd{dominici@fi.infn.it}
\emailAdd{roditi@cbpf.br}

\abstract{We investigate by means of a variational approach the effective potential of a $5D$ $U(1)$ scalar model at finite temperature and compactified on $S^1$  and $S^1/Z_2$ as well as the corresponding $4D$ model obtained through a trivial dimensional reduction. We are particularly interested in the behavior of the thermal masses of the scalar field with respect to the Wilson line phase and the results obtained are compared with those coming from a one-loop effective potential calculation. We also explore the nature of the phase transition.\\
}

\begin{document}

\maketitle
\flushbottom

\noindent

\section{Introduction}

Studies on field theories at finite temperature  have   had a regain of interest particularly  to investigate the electroweak transition and baryogenesis both in the Standard Model (SM) as effective low energy theory  in 4 dimensions
and in its extensions in five dimensions \cite{Grojean:2004xa,Panico:2005ft,Maru:2005jy,Delaunay:2007wb,Hatanaka:2013iya}. These new approaches have some of their theoretical roots in
finite temperature field theory \cite{Matsubara:1955ws,Ezawa:1957rw,Kubo:1957mj,Martin:1959jp}.
Considering topological aspects of a thermal formalism, it is realized that
the prescription results in a scheme of compactification in time of
the $T=0$ theory. That is, the Matsubara results are equivalent to a
path-integral calculated on  $R^{D-1}\times S^{1}$, where $S^{1}$ is a
circle of circumference $\beta =1/T.$ In the case of an extra dimension corresponding to $S^{1}/\mathbf{Z}_2$ orbifold one should also take into account the parity projection  \cite{Kubo:2001zc}. As a consequence, the Matsubara prescription can be thought, in a generalized way, as a mechanism to deal
simultaneously with spatial constraints and thermal effects in a field
theory model. This concept has been developed  for the Matsubara
formalism by considering $R^{D-d}\times S^{l_{1}}\times S^{l_{2}}\cdots \times S^{l_{d}}$, 
with $l_{1}$ corresponding to inverse temperature and $l_{2}\,,\cdots l_{d}$ corresponding to compactification of $d-1$ spatial dimensions. Then the Feynman rules are modified by introducing a generalized Matsubara
prescription, performing the following multiple replacements
(compactification of a $d$-dimensional subspace)~\cite{Khanna:2009zz}, 
\begin{equation}
\int \frac{dk_1}{2\pi }\rightarrow \frac 1{\beta}\sum_{n_1=-\infty }^{+\infty
}\,,\;\;\;\;\int \frac{dk_i}{2\pi }\rightarrow \frac 1{L_i}\sum_{n_i=-\infty }^{+\infty
}\;;\;\;\;k_1\rightarrow \frac{2n_1\pi }{\beta}\;\;\;k_i\rightarrow \frac{2n_i\pi }{L_i}\;,\;\;i=2,3...,d-1,
\label{Matsubara1}
\end{equation}
where $L_i\,,\,\,\;\;i=2,3...,d-1$ are the sizes of the compactified spatial dimensions. 

The ideas above have found new and interesting applications in the context of 
field theories with  extra spatial dimensions, where the Higgs field is identified as the zero mode of the fifth component of the gauge field,  as for instance in~\cite{Fairlie:1979zy,Fairlie:1979at,Manton:1979kb,Hosotani:1983xw,Hosotani:1983vn,Antoniadis:1990ew,Ho:1990xz,Dvali:2001qr,Hall:2001zb,Kubo:2001zc,ArkaniHamed:2001is,Burdman:2002se,Agashe:2004rs,Serone:2005ds,Panico:2005dh,Hosotani:2008tx,Panico:2010is}. 
 Models of this type are sometimes
called models with gauge-Higgs unification.  Perhaps these theories provide an interesting framework
for physics beyond the SM.

In particular, there has been a resurgence of interest in these ideas,
 as a new way to investigate the electroweak
transition and baryogenesis. The electroweak phase
transition, in 5-dimensional finite-temperature field theory with
a compactified extra dimension, has been investigated in~\cite{Panico:2005ft}. 
These authors   compare theirs results with those obtained with the SM at one-loop approximation. The conclusion is for a first-order transition 
with a strength inversely proportional to the Higgs mass, using the estimated value of the Higgs mass,
of $M_{H}<M_{W}/2$. However, these authors also claim that when fermions in the uncompactified 5-D space are introduced, more realistic values of the Higgs mass are obtained. In this case the first-order phase transition becomes weaker. Another interesting result of~\cite{Panico:2005ft} is that up to temperatures of the order 
of $T\approx 1/L$, reliable (low order) perturbative calculations lead to reasonable results. It has been recently noticed that a $SO(5)\times U(1)$ gauge Higgs unification in the Randall Sundrum metric can reproduce the Higgs mass at 125 GeV for three fermion spinorial representations and shown that the thermal phase transition at one loop is first order but very weak so baryogenesis could not occur \cite{Hatanaka:2013iya}.

From a physical and phenomenological point of view, 
an interest in theories with  extra  compactified dimensions at the
inverse TeV scale arose in connection with the new LHC
 experiments. 

In order to go beyond one-loop approximation, self-consistent approaches have been also considered. One example is the Cornwall-Jackiw-Tomboulis effective action for composite operators that has been generalized to finite temperature \cite{Barducci:1986zt,AmelinoCamelia:1992nc,Smet:2001un}. Here we follow an alternative route, which consists in developing another, non perturbative, variational technique, the Gaussian Effective Potential (GEP)~\cite{Stevenson:1984rt,Stevenson:1985zy}, for both finite temperature and compactified spatial dimensions applying it as a tool to investigate effective field theories.

The aim of this paper is to study, by using the GEP, the thermal masses of two models. One being the reduced model, at finite temperature, obtained from neglecting all massive Kaluza Klein modes for a $5$-dimensional $U(1)$ model, thus trivially reducing it to a $4$-dimensional model. The other being the full $5$-dimensional $U(1)$ model, at finite temperature, compactified over $S^{1}$ and over the $S^{1}/\mathbf{Z}_2$ orbifold. Some of our results confirm the perturbative results of~\cite{Panico:2005ft}. 

In Section \ref{reducedmodel} we consider first the GEP for the reduced model where a
truncation of the Kaluza Klein expansion has been performed, so that only the
first KK mode is retained.
In Section \ref{5Dqed} we consider the full 5D scalar electrodynamics, deriving the
GEP at finite temperature and calculating the Higgs thermal mass. We also discuss the structure of the phase transition by looking into the high temperature limit.
 
\section{The reduced model}
\label{reducedmodel}
In this Section we review the reduced model which provides a simple exercise for studying the effective potential for the five dimensional scalar electrodynamics
by truncating the Kaluza Klein tower and considering only a first scalar lower
mode.
Let us consider the following lagrangian:
\be
\LL=(D_\mu \varphi)^\dagger(D^\mu \varphi)+\frac{1}{2}(\partial_\mu s)^2
-g^2 s^2 \varphi^\dagger \varphi -\frac \lambda 4 (\varphi^\dagger \varphi)^2,
\label{lag1}\ee
which corresponds to the reduced $4D$ model.

We introduce real fields ${\hat\phi}_1$, ${\hat\phi}_2$,  
\be
\varphi=\frac 1 {\sqrt{2}}({\hat\phi}_1+i{\hat\phi}_2)
\ee
and perform  a shift on the scalar field $s$,
\be
s(x)=\hat s(x) +s_c
\ee
which allows to interpret $s_c$ as a constant background field.

In the canonically quantized version of  the Gaussian effective potential (GEP) \cite{Stevenson:1984rt,Stevenson:1985zy} the quantum free fields $\hphi$ and $\hat s$, respectively of masses $\Omega$ and $\Delta$, are expanded in the form,   
\be
\hphi=\int \frac{d^3k}{(2\pi)^3 2\omega_k}
\left [ a_\Omega(k)e^{-ikx }
+h.c\right ]
\ee
with
$
\omega_k^2=\vec k^2+\Omega^2
$
and 
\be
\hat s=\int \frac{d^3k}{(2\pi)^3 2\omega_k}
\left [ a_\Delta(k)e^{-ikx }
+h.c\right ]
\ee
with
$
\omega_k^2={\vec k}^2+\Delta^2
$.
Standard commutation relations for creation and annihilation operators are assumed for  $a_\Sigma,~a^\dagger_\Sigma $:
\be
[a_\Sigma(k),a_\Sigma(k')^\dagger ]=2\omega_k (2\pi)^3 \delta^3 (k-k')
\ee
where $\Sigma$ stands for either $\Omega $ or $\Delta$.
In the Gaussian effective potential approach, the effective potential is evaluated
as the expectation value of the Hamiltonian 
\be
V_G(\alpha,\Omega,\Delta)=<0\vert {\cal H}\vert 0>
\label{GEP1}
\ee
where the vacuum $\vert 0\rangle $ is annihilated
by $a_\Omega(k)$ and $a_\Delta(k)$ 
and $\cal H$ is the total Hamiltonian associated to the lagrangian density
(\ref{lag1}). Also we have redefined $s_c=\alpha /gR$, $2\pi R$ being the length of the fifth compactified dimension . One can then minimize $V_G$ with respect to the parameters $\Sigma$ obtaining the associated gap equations which provide the values of the parameters that, when replaced in $V_G $, gives the Gaussian effective potential $\overline V_{G} $

An equivalent calculation may be performed by use of the first order $\delta$-expansion \cite{Okopinska:1987hp,Stancu:1989sk}, which amounts to split the Lagrangian $\cal{L}$ into a sum of a "free field"  ${\cal L}_{0}$ term and a  ${\cal L}_{\delta}$ term containing the interactions, such that when $\delta=1$  the original Lagrangian is recovered. In the next section it will be convenient to cast our calculations into this approach. In both cases, the result for the reduced model is: 
\beq
V_G &=& 2 I_1(\Omega)- \Omega^2 I_0(\Omega)+ I_1(\Delta)-\frac{1}{2}\Delta^2 I_0(\Delta)\label{VG}\\\nonumber
&+& g^2 I_0(\Delta) I_0(\Omega)+\frac{\lambda}{2} I_0^2(\Omega)
+ \frac {\alpha^2}{R^2}I_0(\Omega).
\eeq
where 
\beq
I_0(\Sigma)=\int \frac{d^3p}{(2 \pi)^3}\frac 1 {2\omega_p}\,\,\,\,\,\,\,\,
I_1(\Sigma)=\int \frac{d^3p}{(2 \pi)^3}\frac 1 {2} \omega_p
\eeq
with
$\omega_p=\sqrt{{\bf p}^2+\Sigma^2}$.
The $I_0$ and $I_1$ integrals are equivalent to the covariant form (in the Euclidean and for the $I_1$ case up to an infinite constant)
\be
I_0(\Sigma)=\int \frac{d^4p}{(2 \pi)^4}\frac 1 {p^2+\Sigma^2}\,\,\,\,\,
I_1(\Sigma)=\frac 1 2\int \frac{d^4p}{(2 \pi)^4}\log{[p^2+\Sigma^2]}
\ee

The gap equations are:
\beq
\Omega^2 &=&\frac {\alpha^2}{R^2} + \lambda I_0(\Omega) 
+g^2 I_0(\Delta) ,\label{Delta}\\\nonumber
\Delta^2 &=& 2 g^2 I_0(\Omega).
\eeq
where use has been done of the identity 
\beq
\frac d {d\Sigma}I_1(\Sigma)=\Sigma I_0(\Sigma)
\label{derivee}
\eeq

Replacing the values of $\Omega$ and $\Delta$ from Eq.(\ref{Delta}) into Eq.(\ref{VG}) gives the 'optimized' Gaussian effective potential, $\overline V_{G}.$
The divergent integrals $I_0,I_1$ are evaluated using dimensional
regularization with the minimal subtraction scheme ($\overline {\rm{MS}}$) prescription.
In particular one has (in the Euclidean and using their covariant form)
\beq
I_0(\Sigma)=\int \frac {d^4p}{(2\pi)^4}\frac 1 {p^2+\Sigma^2}=
\frac {\Sigma^2} {(4\pi)^2}\left (\log{\frac {\Sigma^2} {\mu^2}}-1\right )
\eeq
\beq
I_1(\Sigma)=\frac 1 2 \int \frac {d^4p}{(2\pi)^4}\log (p^2+\Sigma^2)=
\frac {\Sigma^4} {64\pi^2}\left (
\log{\frac {\Sigma^2} {\mu^2}}-\frac 3 2\right )\eeq
where $\mu$ is a regularization scale.
Using this regularization prescription,
Eq.(\ref{derivee})
is still valid and therefore the form of the gap equation is the same as
Eq.(\ref{Delta}).

The finite temperature result can be obtained by replacing the $I$ integrals by their finite temperature version $I^{FT},$ as described in \cite{Roditi:1986tu,Hajj:1987gk}. For $I_{1}^{FT}$, one has
\be
I_{1}^{FT} = I_{1} + I_{1}^{\beta}
\ee
where the parameter $\beta$ indicates, as usual, the inverse of the temperature $T$ and
\be
I_{1}^{\beta}(\Sigma)=\frac{1}{\beta} \int \frac {d^3 p}{(2\pi)^3}\log (1- e^{-\beta w_p})=
\frac 1 {\beta^4}\left [ -\frac {\pi^2}{90}+\frac {(\beta\Omega)^2}
 {24}-\frac {(\beta\Omega)^3}
 {12\pi} -\frac {(\beta\Omega)^4}
 {64\pi^2}(\log\beta^2\Omega^2-c)\right ], 
\ee
with $c=3/2+ 2(\log 4\pi-\gamma)\sim 5.41$.
It is important to observe that the structure of the Gaussian effective potential is maintained as Eq.(\ref{derivee}) is also valid for the finite temperature integral $I_{1}^{FT}.$ With these substitutions one is then led to a finite temperature Gaussian effective potential ${\overline V}_{G}^{FT}.$

Having obtained the optimized finite temperature Gaussian effective potential we can calculate the Wilson line thermal mass of the field $\alpha$, i.e.,  

\be
M^2(T,\alpha)={\cal N} \frac{\partial^2 {\overline V_{G}^{FT}}}{\partial \alpha^2},
\ee
where $\cal N$ is a suitable normalization factor.

\begin{figure}
\begin{center}
\includegraphics[width=3in]{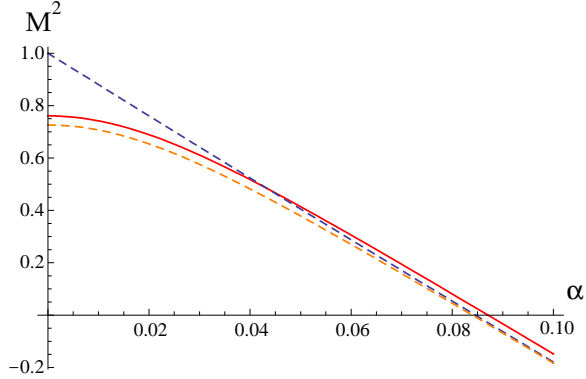}
\caption{Squared thermal mass of the field $\alpha$ in the reduced model from the GEP (continuous line),
one-loop effective potential (upper dashed line), improved one-loop (lower dashed line)
 for $LT=1$, $\lambda=1$, $\mu=1/R$.}
\label{Fig1}
\end{center}
\end{figure}

As we are dealing with an effective theory, a series of approximations are in order and so, following \cite{Panico:2005ft} we shall assume that $g$ is negligible. In that case, we can consider that, from Eq.(\ref{Delta}), $\Delta=0$, and within the dimensional regularization scheme one can set scale-independent integrals as $I_0(0)$ or $I_1(0)$ equal to zero. So we are led to a simpler situation where only the equation for $\Omega$ needs to be satisfied. Assuming that $LT=1$, $\lambda = 1$ and $\mu=1/R$, the scale above which one considers that the effective theory breaks down, we plot $M^2(T,\alpha)$, in Fig.\ref{Fig1}, for  ${\overline V}_{G}^{FT}$ (red line) and compare it with the results for the improved one-loop effective potential (red dashed line) and the
"{\i}naive one-loop effective potential (upper dashed line) taken from \cite{Panico:2005ft}, which, in our notation, is simply proportional to the integral $I_1^{FT}(\Omega)$ for a value $\Omega=\alpha/R$. The normalization factor $\cal N$ is chosen such as to make equal to $1$ the thermal mass calculated from the na\"{\i}ve one-loop effective potential when $\alpha = 0.$
The main result is that the thermal mass calculated with GEP is substantially in agreement (slightly higher)
with  the prediction of the improved one-loop method. 

\section{Compactified 5D scalar QED}
\label{5Dqed}

Let us now consider  5D $U(1)$ scalar electrodynamics, with the 5th dimension compactified
on a circle $S^1$ of length $L=2\pi R$, defined in a Euclidean space by the
action 
\begin{eqnarray}
  S_E&=& \int_{-\pi R}^{\pi R} dy\int d^4x  \Big [\frac 1 4 F_{MN}F_{MN}+(D_M\varphi)^\dagger(D_M\varphi)
\nn\\
&+& \frac {\lambda_5}{4} (\varphi^\dagger\varphi)^2+{\cal L}_{gf}
 \Big ]
\end{eqnarray}
with ${\cal L}_{gf}$ being the gauge fixing term and
\be
D_M \varphi=\partial_M \varphi +i g_5 A_M\varphi ,
\ee
where $\varphi$ satisfies the boundary condition $\varphi(2\pi R)=\varphi(0)$
 and capital letters indicate $5D$ indices.
Notice that the change from Minkowskian to Euclidean space corresponds to $V_MT^M\to -V_MT_M$, therefore after
a redefinition of sign in the action. We adopt the convention that capital Roman letter label indices running from $1$ to $5$, while Greek letters are used for indices which run from $1$ to $4$. 
We are interested in considering the effective potential for the vacuum solution for the 
5th component of the field $A_M$. Therefore we will consider
the $A_M$ field as a background field with only $A_5({\bf x},x_5)\neq 0$, where we have used the notation ${\bf x}\equiv (x_1,\cdots,x_4)$ for the coordinates of the position vector in the non-compactified subspace. Here, as mentioned, it will be more convenient to introduce the Gaussian effective potential as the first order in the $\delta-$expansion \cite{Okopinska:1987hp,  Stancu:1989sk}.

Introducing the real components of $\varphi$,
\be
\varphi=\frac 1 {\sqrt{2}}(\hat\phi_1+i\hat\phi_2), 
\ee
the  part of the  Lagrangian quadratic in ${\hat\phi}_{1,2}$ can be rewritten as
\be
\frac 1 2 \Phi^T D^{-1} \Phi
\ee
with $\Phi=(\hat\phi_1,\hat\phi_2)$
\be
D^{-1}=\left(
\begin{array}{cc}
-\partial_M^2+g_5^2 A_5^2&2 g_5 A_M\partial_M\\
-2 g_5 A_M\partial_M& -\partial_M^2+g_5^2 A_5^2\\
\end{array}
\right)
\ee
Then we shift the field $A_5$ by a constant background field ${\cal A}_5$ getting,
\be
\hat A_5({\bf x},x_5)=A_5({\bf x},x_5)-{\cal A}_5
\ee
and  we split the lagrangian as
\be
\LL=(\LL_0+\LL_{int})_{\delta=1}
\ee
where $\LL_0$ 
is the sum of the quadratic lagrangian for $\hat \phi_{1,2}$, the electromagnetic field terms and the terms containing the variational parameter $\Omega$:
\be
\LL_0=
\frac 1 2\left [\hPhi \hat {\cal D} \hPhi+(\partial_\mu {\hat A_5})^2\right ]
\ee
with
\be
\hat {\cal D}^{-1} =\left(
\begin{array}{cc}
-\partial_M^2+\frac{\alpha^2}{R^2}+\Omega^2
&2 \frac{\alpha}{R}\partial_5\\
-2 \frac{\alpha}{R}\partial_5& -\partial_M^2+\frac{\alpha^2}{R^2}+ \Omega^2\\
\end{array}
\right),
\ee
where we have used $g_5{\cal A}_5={\alpha}/{R}$ as well as $\lambda_5=L\lambda.$

The interaction Lagrangian is then:
\beq
{\cal L}_{int} & = &
\delta\Bigg \{ \frac{1}{2}\left[\frac{\alpha^2}{R^2}-\Omega^2\right]\left(\hat{\phi}_{1}^{2}+\hat{\phi}_{2}^{2}\right)
+ \frac {L\lambda}{16} \hat{\phi}_{1}^{4}
+  \frac {L\lambda}{16}\hat{\phi}_{2}^4+ \frac {L\lambda}{8}\hat{\phi}_{1}^2 \hat{\phi}_{2}^2
\nonumber\\
&+& \frac{1}{2}g_{5}^2
\hat A_{5}^2\left(\hat{\phi}_{1}^2+\hat{\phi}_{2}^2\right) - \Delta^2 \hat A_{5}^2
\Bigg\}.
\label{lint}
\eeq
In Eq.(\ref{lint}), for simplicity, we kept only the quadratic terms that survive after the functional integrations.

Introducing sources for the fields, the generating functional
is written as:
\be
\label{zj}
Z[J,j]=\int \mathrm{D}\left[\hat\phi_1,\hat\phi_2,\hat A_{5} \right] \;
\exp \left(\mbox{}-\int_{-\pi R}^{\pi R} d{x}_5\int d^4x\,\{\LL_0+\LL_{int}\} \,+ \, \langle J.\Phi \rangle \,+\, \langle j.A_5({\bf x},x_5)\rangle \right),
\ee
and the effective action is obtained by the Legendre transformation
\be
\Gamma[\alpha]=\ln Z[j] \,- \langle j.A_5\rangle,
\label{action1}
\ee
where the brackets $\langle\,\rangle$ are a shorthand notation for the integral 
$\int dx_5\,d^{4}x.$

One can expand the action, Eq.(\ref{action1}), to first order in $\delta$ by the usual functional techniques getting,  
\be
 \Gamma_{GEP}= \frac{1}{2} Tr~ ln~ ({\cal D}^{-1})  + \frac{1}{2} Tr~ ln~ (G_{5}^{-1}) + \delta <\LL(\frac{\delta}{\delta (J,j)})> exp[\frac 1 2 (<J^{T}.{\cal D}.J>+<j G_{5} j>)]{\bigg |}_{J,j=0}
 \label{action2}
 \ee
where, again, the brackets are a short notation for the integrations and the summations over the compactified variables, $\LL(\frac{\delta}{\delta (J,j)})$ means that we have replaced the fields $\hat\phi_1,\hat\phi_2$ and ${\hat A_5}$ by their functional derivatives in $J_1,J_2$ and $j$. The inverse of $\hat {\cal D}$ in the momentum representation is given by
\be
\hat {\cal D}^{-1}=
\left(
\begin{array}{cc}
p_M^2+\frac{\alpha^2}{R^2}+\Omega^2
&2 i\frac{\alpha}{R}p_5\\
-2 i\frac{\alpha}{R}p_5& p_M^2+\frac{\alpha^2}{R^2}+ \Omega^2\\
\end{array}
\right).
\ee

The result, after the calculation of the traces and omitting the terms in g (as mentioned at the end of the previous section, $g$ is considered negligible), reduces to
\be
V_{G}= {\tilde{I}}_1(\Omega)-\frac 1 2\Omega^2{\tilde{I}}_0(\Omega) + \frac 1 2\frac{{\alpha}^2}{R^2}{\tilde{I}}_0(\Omega)+\frac{L\lambda}{2}{\tilde{I}}^2_0(\Omega), 
\label{Vgep5}
\ee
where the integrals $I_0$ and $I_1$, prior to compactification are given by:
\beq
{\tilde I}_0(\Omega)&=&\left [ \int\!\frac{d^5p}{(2\pi)^5}\,\frac 1 {p^2 + (p_5 + \frac{\alpha}{R})^2+\Omega^2}+
\int\!\frac{d^5p}{(2\pi)^5}\,\frac 1 {p^2 + (p_5 - \frac{\alpha}{R})^2+\Omega^2}
\right ] \nonumber\\ &=& \sum_{q=\pm1} \int\!\frac{d^5p}{(2\pi)^5}\,\frac 1 {p^2 + (p_5 + q \frac{\alpha}{R})^2+\Omega^2}
\label{I0}
\eeq
and
\beq
{\tilde{I}}_1(\Omega)&=& \frac 1 2\int\!\frac{d^5p}{(2\pi)^5}\, \mathrm{log}\left[\left(p^2 + (p_5 + \frac{\alpha}{R})^2+\Omega^2\right)\left(p^2 + (p_5 - \frac{\alpha}{R})^2+\Omega^2\right)\right]\nonumber\\ &=& \sum_{q=\pm1} \frac 1 2\int\!\frac{d^5p}{(2\pi)^5}\, \mathrm{log}\left[\left(p^2 + (p_5 + q\frac{\alpha}{R})^2+\Omega^2\right)\right].
\eeq
The gap equation, obtained by minimizing the potential $V_G$ with respect to the parameter $\Omega$, in this case is:
\be
\Omega^2 =\frac {\alpha^2}{R^2} + 2 L\lambda {\tilde I}_0(\Omega) .
\label{gap5d}
\ee

As before replacing the value of $\Omega$ Eq.(\ref{gap5d}) into Eq.(\ref{Vgep5}) gives the value of Gaussian effective potential, $\overline V_{G}$, now for the compactified scalar QED. Formally, when the value $\overline{\Omega}$ gives a minimum, it is possible to replace Eq.(\ref{gap5d}) in Eq.(\ref{Vgep5}), such that:
\be
\overline V_{G}={\tilde{I}}_1(\overline{\Omega})-\frac{L\lambda}{2}{\tilde{I}}^2_0(\overline{\Omega}).\label{Vmin}
\ee

In order to introduce simultaneously the compactification of the $5$th dimension over $S^1$ and a finite temperature, one has to perform 
the replacements,
\begin{equation}
\frac{d^5p}{(2\pi)^5}\rightarrow \frac{T}{L}\sum_{k,n=-\infty}^{+\infty} \int\!\frac{d^3p}{(2\pi)^3}\,;\;\;\;\;\;p_5 \rightarrow \frac{2\pi k}{L}\,,\;\;\;\;\;p_4 \rightarrow 2\pi nT.
\label{matsubara}
\end{equation}
Remembering that $L=2\pi R$, we finally get, 
\be
{\tilde{I}}_0^{FT}(\Omega)= \frac{T}{4\pi R}\sum_{q=\pm1} \sum_{k,n=-\infty}^{+\infty} \int\!\frac{d^3p}{(2\pi)^3}
\frac {1}{ \bigg[{\vec p}^2 + \bigg(2\pi n T\bigg)^2 + \Omega^2 + \frac{\dd (k + q\alpha)^2}{\dd R^2}\bigg]  }
\label{I05d}
\ee
and
\be
{\tilde{I}}_1^{FT}(\Omega)= \frac{T}{8\pi R}\sum_{q=\pm1} \sum_{k,n=-\infty}^{+\infty} \int\!\frac{d^3p}{(2\pi)^3}
\log \bigg[ {\vec p}^2 + \bigg(2\pi n T\bigg)^2 + \Omega^2 + \frac{\dd(k + q\alpha)^2}{\dd R^2}\bigg] \,.
\label{I15d}
\ee
In Eqs.~(\ref{matsubara}) to (\ref{I15d}) the sums over $n$ and $k$ refer respectively to the Matsubara and compactification modes. 
The motivation to go through this procedure, in order to obtain Eq.(\ref{I15d}), is that this expression is much easier to manipulate by means of the Poisson summation formula,
\be
\sum_{\ell=-\infty}^{+\infty}\,e^{-\pi t(\ell+a)^{2}}\,=\,\frac{1}{\sqrt{t}}\sum_{\kappa=-\infty}^{+\infty}\,e^{-\frac{\pi \kappa^2}{t}} e^{2i\pi \kappa a}\,=\,\frac{1}{\sqrt{t}}\left[1+2\sum_{\kappa=1}^{+\infty}\,e^{-\frac{\pi \kappa^2}{t}}cos(2\pi \kappa a)\right],
\label{poisson}
\ee
together with the identity 
\be
\textrm{tr}\,log M = -\int_{0}^{\infty} \frac{dt}{t}\, \textrm{tr} \, e^{-tM}.
\label{tidn}
\ee

Using Eqs.(\ref{poisson}), (\ref{tidn}) and the representation of  Bessel functions 
of the third kind, $K_{\nu }$,
\begin{equation}
2(A/B)^{\frac{\nu }{2}}K_{\nu }(2\sqrt{AB})=\int_{0}^{\infty }dx\;x^{-\nu
-1}e^{-(A/x)-Bx},
\label{K}
\end{equation}
the integrals,  ${\tilde{I}}_0^{FT}$ and ${\tilde{I}}_1^{FT}$, subtracting the divergent term
corresponding to the zero mode, which however does not depend on
$\alpha$~\cite{Panico:2005ft}, can be written as:

\begin{equation}
{\tilde{I}}_0^{FT}=\,\frac{T^3} {2(2\pi)^{5/2}}\sum_{q=\pm1} \sum_{\kappa=1}^{+\infty} \sum_{\ell=-\infty}^{+\infty} \frac{cos(2\pi \kappa q \alpha)}{[\ell^{2} + (LT\kappa)^{2}]^{\frac{3}{2}}} B_{\frac{3}{2}}(\frac{\Omega}{T}\sqrt{\ell^{2} + (LT\kappa)^{2}})
\label{ireg0}
\end{equation}
and 

\begin{equation}
{\tilde{I}}_1^{FT}=-\frac{2T^5} {(2\pi)^{5/2}}\sum_{q=\pm1} \sum_{\kappa=1}^{+\infty} \sum_{\ell=-\infty}^{+\infty} \frac{cos(2\pi \kappa q \alpha)}{[\ell^{2} + (LT\kappa)^{2}]^{\frac{5}{2}}} B_{\frac{5}{2}}(\frac{\Omega}{T}\sqrt{\ell^{2} + (LT\kappa)^{2}}),
\label{ireg1}
\end{equation}
where $B_{\nu}(z)=z^{\nu}K_{\nu }(z)$. 

In the calculations above we have performed the Poisson resummation over both the Matsubara ($\ell$) and compactification ($\kappa$) modes. Equivalently, it is possible to use this procedure only over the Matsubara modes or only over the compactification ones. The results are:

\begin{equation}
{\tilde{I}}_0^{FT}=\,\frac{T} {8(2\pi)^{2}L^2} \sum_{q=\pm1}\sum_{\kappa=1}^{+\infty} \sum_{\ell=-\infty}^{+\infty} \frac{cos(2\pi \kappa q \alpha)}{ \kappa^{2}} B_{1}(L\kappa\sqrt{\Omega^2 + (2\pi T \ell)^{2} }),
\label{imats0}
\end{equation}
and 
\begin{equation}
{\tilde{I}}_1^{FT}=-\frac{2T} {(2\pi)^{2}L^4}\sum_{q=\pm1} \sum_{\kappa=1}^{+\infty} \sum_{\ell=-\infty}^{+\infty} \frac{cos(2\pi \kappa q \alpha)}{ \kappa^{4}} B_{2}(L\kappa\sqrt{\Omega^2 + (2\pi T \ell)^{2} }),
\label{imats1}
\end{equation}
for the compactification modes, or
\beq
{\tilde{I}}_0^{FT}&=&\,\frac{1} {(2\pi)^{5/2}L^3} \sum_{q=\pm1} \sum_{\kappa=1}^{+\infty} \frac{cos(2\pi \kappa q \alpha)}{ \kappa^{3}} B_{3/2}(L\Omega\kappa)\nonumber\\
&+&\frac{T^2}{(2\pi)^2L} \sum_{q=\pm1}\sum_{\kappa=-\infty}^{+\infty} \sum_{\ell=1}^{+\infty}\frac{1}{\ell^2}  B_1(\frac{\ell}{T}\sqrt{\Omega^2 +\frac{1}{R^2}(\kappa +q\alpha)^2})
\label{icomp1}
\eeq
and 
\beq
{\tilde{I}}_1^{FT}&=&-\frac{2} {(2\pi)^{5/2}L^5} \sum_{q=\pm1} \sum_{\kappa=1}^{+\infty} \frac{cos(2\pi \kappa q \alpha)}{ \kappa^{5}} B_{5/2}(L\Omega\kappa)\nonumber\\
&-&\frac{2T^4}{(2\pi)^2L} \sum_{q=\pm1}\sum_{\kappa=-\infty}^{+\infty} \sum_{\ell=1}^{+\infty}\frac{1}{\ell^4} B_2(\frac{\ell}{T}\sqrt{\Omega^2 +\frac{1}{R^2}(\kappa +q\alpha)^2})
\label{icomp2}
\eeq
for the Matsubara modes.

\begin{figure}
\begin{center}
\includegraphics[width=3in]{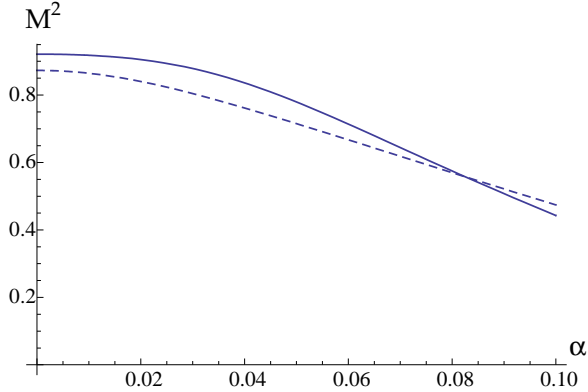}
\caption{Squared thermal mass of the field $\alpha$ in the 5D  model using the GEP (continuous line) and the
 improved one-loop (dashed line) for $LT=1$ and $\lambda=1$.}
\label{Fig2}
\end{center}
\end{figure}

We have then laid the setup for the calculation of the thermal mass of the field $\alpha$.
In Fig.\ref{Fig2} we compare the results  of the squared thermal mass $M^2$ obtained with the GEP (red continuous line), for  $LT=1$  and $\lambda=1$, with the improved one-loop effective potential \cite{Panico:2005ft}. The variational calculation result is larger than the improved one-loop 
for small values of $\alpha$ and smaller for larger value. In fact, the variational calculation interpolates the naif one-loop and improved results for the region of small values of $\alpha$ and is more sensitive than the perturbative approaches in the larger values of $\alpha$ region. 

It is also interesting to notice that, as in \cite{Panico:2005ft}  we can obtain an indication of the order of the phase transition. Using the expressions of  eqs.(\ref{imats0}) and (\ref{imats1}) for the end point $\Omega=0$ and dropping all the modes, but the zero mode, in the Matsubara expansion we arrive at:
\be
{\tilde{I}}_0^{FT} =\,\frac{T} {8(2\pi)^{2}L^2} \sum_{q=\pm1} \sum_{\kappa=1}^{+\infty} \frac{cos(2\pi \kappa q \alpha)}{ \kappa^{2}}
\label{i0zeta}
\ee
and
\be
{\tilde{I}}_1^{FT} =-\frac{T} {\pi^2L^4} \sum_{q=\pm1}\sum_{\kappa=1}^{+\infty} \frac{cos(2\pi \kappa q \alpha)}{ \kappa^{4}},
\label{i1zeta}
\ee
where we have also used ${\rm lim}_{\xi \to 0}\,B_{\nu}(\xi)=2^{\nu - 1}\Gamma(\nu).$

The  cosinus function in Eqs.(\ref{i0zeta}) and (\ref{i1zeta}) lead to  ill-defined, rapidly oscillating series. 
Nevertheless, there is a solution  in the framework of the $zeta$-function regularization. This 
procedure has already been used in~\cite{Panico:2005ft} 
and it is a crucial step to these authors conclude for a first-order electroweak transition.  As it is well-known, 
the zeta function can be analytically extended to the whole complex plane, having 
only one pole at $z=1$. This analytical extension with a\textit{\
strictly negative} \textit{even }argument vanishes: $\zeta (-2n)=0$ for
integer $n\geq 1$. 
Accordingly, we expand the cosinus functions above  in a power series of $\alpha$ and 
using 
$\sum_{\kappa=1}^{+\infty}\kappa^{2n}=\zeta(-2n)=0$ for all positive integers $n\geq 1$ it is possible to get, for definite values of $q$ (we use $\zeta(0)=-1/2)$):
\be
{\tilde{I}}_0^{FT} =\frac{T}{32 L^2} (\frac{1}{6} - q\alpha + (q\alpha)^2)
\label{i0zreg}
\ee
and 
\be
{\tilde{I}}_1^{FT} = -\frac{\pi^2 T}{3 L^4}(\frac{1}{30} - (q\alpha)^2 + 2(q\alpha)^3 - (q\alpha)^4)
\label{i1zreg}
\ee

Now, the reasoning for doing these manipulations is that the $\Omega=0$ endpoint corresponds to an infrared limit \cite{Stevenson:1984rt, Panico:2005ft}, also one can see that in eqs.(\ref{imats0}) and (\ref{imats1}) when the temperature, $T$, is very large and for any non null  $\ell$ , the generalized Bessel function $B_{\lambda}$ approaches zero and  the extremum values for the integrals occurs at the $\Omega=0$ end point.
Under these conditions the potential will be given by replacing eqs. (\ref{i0zreg}) and (\ref{i1zreg}) in Eq.(\ref{Vgep5}) taking the endpoint $\Omega=0$.

In the $S^1$ compactification case we keep all degrees of freedom and  the odd powers of $\alpha$ cancel out . One gets, apart from a term independent of $\alpha$, a polynomial of order $4$ in $\alpha$ for the potential, which has the form,
\begin{equation}
{V}_G = \frac {\pi^2}{48 L^5}(A\alpha^2 + B\alpha^4)
\label{alpha4}
\end{equation} 
 where the coefficients $A =33 \rho +\lambda \rho^2/(32 \pi^2)$ and $B =38 \rho +3 \lambda \rho^2/(32 \pi^2) $ are positive and $\rho=LT$. 
 Since we are in the high temperature regime $TL>>1$, we expect the system to be in the disordered phase.
 So, for $\lambda$ positive and when we keep all degrees of freedom, the form of Eq.(\ref{alpha4}) indicates the system has undergone a second-order phase transition.  

 The $S^1/ Z_2$ compactification runs in a similar way, with the caveat that one has to project the states over a definite value for the charge $q$~\cite{Kubo:2001zc},  we have to choose only one value of $q$ for the expressions in Eqs.~(\ref{i0zreg}) and (\ref{i1zreg}); in this case the odd powers of $\alpha$ do not cancel. That means that an $\alpha^3$ term remains in the effective potential. Its form is proportional to $\alpha^3 (35/48 \rho + \lambda \rho^2/(32\pi)^2$
 indicating that for  positive values of $\lambda$ the system has undergone a first order phase transition.

\acknowledgments I.R. thanks the warm hospitality of the Dipartimento di Fisica Universit\`a di Firenze where parts of this work were done. He also acknowledges fruitful discussions as well as crucial suggestions from Adolfo P.C. Malbouisson. The authors acknowledge partial financial support from INFN,  Compaq and PRIN contract 2010YJ2NYW.



\begin{thebibliography}{10}

\bibitem{Grojean:2004xa}
C.~Grojean, G.~Servant and J.~D. Wells,
\newblock Phys. Rev. {\bf D71}, 036001 (2005), [hep-ph/0407019].

\bibitem{Panico:2005ft}
G.~Panico and M.~Serone,
\newblock JHEP {\bf 05}, 024 (2005), [hep-ph/0502255].

\bibitem{Maru:2005jy}
N.~Maru and K.~Takenaga,
\newblock Phys. Rev. {\bf D72}, 046003 (2005), [hep-th/0505066].

\bibitem{Delaunay:2007wb}
C.~Delaunay, C.~Grojean and J.~D. Wells,
\newblock JHEP {\bf 04}, 029 (2008), [0711.2511].

\bibitem{Hatanaka:2013iya}
H.~Hatanaka,
\newblock 1304.5104.

\bibitem{Matsubara:1955ws}
T.~Matsubara,
\newblock Prog. Theor. Phys. {\bf 14}, 351 (1955).

\bibitem{Ezawa:1957rw}
H.~Ezawa, Y.~Tomozawa and H.~Umezawa,
\newblock Nuovo Cim. {\bf 5}, 810 (1957).

\bibitem{Kubo:1957mj}
R.~Kubo,
\newblock J. Phys. Soc. Jap. {\bf 12}, 570 (1957).

\bibitem{Martin:1959jp}
P.~C. Martin and J.~S. Schwinger,
\newblock Phys. Rev. {\bf 115}, 1342 (1959).

\bibitem{Kubo:2001zc}
M.~Kubo, C.~Lim and H.~Yamashita,
\newblock Mod.Phys.Lett. {\bf A17}, 2249 (2002), [hep-ph/0111327].

\bibitem{Khanna:2009zz}
F.~C. Khanna, A.~P. Malbouisson, J.~M. Malbouisson and A.~R. Santana,
\newblock (2009),
\newblock World Scientific, New Jersey, 2009 (ISBN-13: 978-981-281-887-4,
  ISBN-10: 981-281-887-1, ebook ISBN-13: 978-981-281-889-8, ebook ISBN-10:
  981-281-889-8).

\bibitem{Fairlie:1979zy}
D.~Fairlie,
\newblock J.Phys.G {\bf G5}, L55 (1979).

\bibitem{Fairlie:1979at}
D.~Fairlie,
\newblock Phys.Lett. {\bf B82}, 97 (1979).

\bibitem{Manton:1979kb}
N.~Manton,
\newblock Nucl.Phys. {\bf B158}, 141 (1979).

\bibitem{Hosotani:1983xw}
Y.~Hosotani,
\newblock Phys.Lett. {\bf B126}, 309 (1983).

\bibitem{Hosotani:1983vn}
Y.~Hosotani,
\newblock Phys.Lett. {\bf B129}, 193 (1983).

\bibitem{Antoniadis:1990ew}
I.~Antoniadis,
\newblock Phys.Lett. {\bf B246}, 377 (1990).

\bibitem{Ho:1990xz}
C.-L. Ho and Y.~Hosotani,
\newblock Nucl. Phys. {\bf B345}, 445 (1990).

\bibitem{Dvali:2001qr}
G.~Dvali, S.~Randjbar-Daemi and R.~Tabbash,
\newblock Phys.Rev. {\bf D65}, 064021 (2002), [hep-ph/0102307].

\bibitem{Hall:2001zb}
L.~J. Hall, Y.~Nomura and D.~Tucker-Smith,
\newblock Nucl.Phys. {\bf B639}, 307 (2002), [hep-ph/0107331].

\bibitem{ArkaniHamed:2001is}
N.~Arkani-Hamed, A.~G. Cohen and H.~Georgi,
\newblock Phys.Lett. {\bf B516}, 395 (2001), [hep-th/0103135].

\bibitem{Burdman:2002se}
G.~Burdman and Y.~Nomura,
\newblock Nucl.Phys. {\bf B656}, 3 (2003), [hep-ph/0210257].

\bibitem{Agashe:2004rs}
K.~Agashe, R.~Contino and A.~Pomarol,
\newblock Nucl.Phys. {\bf B719}, 165 (2005), [hep-ph/0412089].

\bibitem{Serone:2005ds}
M.~Serone,
\newblock AIP Conf.Proc. {\bf 794}, 139 (2005), [hep-ph/0508019].

\bibitem{Panico:2005dh}
G.~Panico, M.~Serone and A.~Wulzer,
\newblock Nucl.Phys. {\bf B739}, 186 (2006), [hep-ph/0510373].

\bibitem{Hosotani:2008tx}
Y.~Hosotani, K.~Oda, T.~Ohnuma and Y.~Sakamura,
\newblock Phys.Rev. {\bf D78}, 096002 (2008), [0806.0480].

\bibitem{Panico:2010is}
G.~Panico, M.~Safari and M.~Serone,
\newblock JHEP {\bf 1102}, 103 (2011), [1012.2875].

\bibitem{Barducci:1986zt}
A.~Barducci, R.~Casalbuoni, D.~Dominici, R.~Gatto and G.~Pettini,
\newblock Phys. Lett. {\bf B179}, 275 (1986).

\bibitem{AmelinoCamelia:1992nc}
G.~Amelino-Camelia and S.-Y. Pi,
\newblock Phys. Rev. {\bf D47}, 2356 (1993), [hep-ph/9211211].

\bibitem{Smet:2001un}
G.~Smet, T.~Vanzielighem, K.~Van~Acoleyen and H.~Verschelde,
\newblock Phys. Rev. {\bf D65}, 045015 (2002), [hep-th/0108163].

\bibitem{Stevenson:1984rt}
P.~M. Stevenson,
\newblock Phys. Rev. {\bf D30}, 1712 (1984).

\bibitem{Stevenson:1985zy}
P.~M. Stevenson,
\newblock Phys. Rev. {\bf D32}, 1389 (1985).

\bibitem{Okopinska:1987hp}
A.~Okopinska,
\newblock Phys. Rev. {\bf D35}, 1835 (1987).

\bibitem{Stancu:1989sk}
I.~Stancu and P.~M. Stevenson,
\newblock Phys. Rev. {\bf D42}, 2710 (1990).

\bibitem{Roditi:1986tu}
I.~Roditi,
\newblock Phys. Lett. {\bf B169}, 264 (1986).

\bibitem{Hajj:1987gk}
G.~A. Hajj and P.~M. Stevenson,
\newblock Phys. Rev. {\bf D37}, 413 (1988).

\end{thebibliography}

\end{document}